\documentstyle[12pt,aaspp4]{article}

\def\lsim{\lower.5ex\hbox{$\; \buildrel < \over \sim \;$}}
\def\gsim{\lower.5ex\hbox{$\; \buildrel > \over \sim \;$}}

\def\t{\ifmmode {\tau} \else $\tau$ \fi}

\def\ref{\noindent \hangafter=1 \hangindent=0.7 truecm}

\def\cm{\ifmmode {\rm cm}^{-1} \else cm$^{-1}$ \fi}
\def\s{\ifmmode {\rm s}^{-1} \else s$^{-1}$ \fi}
\def\cc{\ifmmode {\rm cm}^{-3} \else cm$^{-3}$ \fi}
\def\cs{\ifmmode {\rm cm}^{-2} \else cm$^{-2}$ \fi}
\def\g{\ifmmode \gamma \else $\gamma$\fi}
\def\G{\ifmmode \Gamma \else $\Gamma$\fi}

\def\kms{\ifmmode {\rm km\ s}^{-1} \else km s$^{-1}$\fi}

\tighten

\begin{document}

\title{Optical and Radio Polarimetry of the M87 Jet at $0.2''$ Resolution}

\author{Eric S. Perlman, John A. Biretta, Fang Zhou, William B. Sparks, and 
F. Duccio Macchetto}

\affil{Space Telescope Science Institute, 3700 San Martin Drive, Baltimore,
MD  21218}

\begin{abstract}

We discuss optical (HST/WFPC2 F555W) and radio (15 GHz VLA)
polarimetry observations of the M87 jet taken during 1994-1995.  The
angular resolution of both of these observations is $\sim 0.2''$,
which at the distance of M87 corresponds to 15 pc.  Many knot regions
are very highly polarized ($\sim 40-50\%$, approaching the theoretical
maximum for optically thin synchrotron radiation), suggesting highly
ordered magnetic fields.  High degrees of polarization are also
observed in interknot regions.  The optical and radio polarization
maps share many similarities, and in both, the magnetic field is
largely parallel to the jet, except in the ``shock-like'' knot regions
(parts of HST-1, A, and C), where it becomes perpendicular to the jet.

We do observe significant differences between the radio and optical
polarized structures, particularly for bright knots in the inner jet,
giving us important insight into the radial structure of the jet.
Unlike in the radio, the optical magnetic field position angle becomes
perpendicular to the jet at the upstream ends of knots HST-1, D, E and
F.  Moreover, the optical polarization appears to decrease markedly at
the position of the flux maxima in these knots.  In contrast, the
magnetic field position angle observed in the radio remains parallel
to the jet in most of these regions, and the decreases in radio
polarization are smaller.  More minor differences are seen in other
jet regions.  Many of the differences between optical and radio
polarimetry results can be explained in terms of a model whereby
shocks occur in the jet interior, where higher-energy electrons are
concentrated and dominate both polarized and unpolarized emissions in
the optical, while the radio maps show strong contributions from
lower-energy electrons in regions with {\bf B} parallel, near the jet
surface.

\end{abstract}


\vfill\eject

\section{Introduction}

The synchrotron jet hosted by the giant elliptical galaxy M87 is
perhaps the most intensely studied feature of any active galaxy (AGN).
Due to its brightness and proximity (distance 16 Mpc; Tonry 1991), it
represents an ideal testing ground for jet models.  Its large-scale
radio structure (Biretta 1993, B\"ohringer et al. 1995) is typical of
low-power FR 1 radio galaxies (Fanaroff and Riley 1974): the radio
lobes are edge-dimmed, and a prominent, apparently one-sided jet
emerges from the nucleus.  No counterjet has been seen in any optical
or radio image of the jet; however, there is strong evidence for its
presence in the form of optical synchrotron emission at the location
of a radio hotspot in the southeast lobe (Sparks et al. 1992,
Stiavelli et al. 1992).

While the gross morphology of the M87 jet is remarkably constant from
the radio through the optical (Boksenberg et al. 1992), small but
significant differences have been found at the highest resolutions by
pre- and post-COSTAR {\it HST} observations (Sparks, Biretta \&
Macchetto 1996, hereafter SBM96).  In particular, the optical jet is
considerably more compact than the radio jet, and the knots are more
centrally concentrated.  These trends appear to continue from the
optical into the UV.  Recent X-ray images suggest that more
differences may be present in the ROSAT band (Neumann et al. 1997).
The jet is highly polarized (Baade 1956, Warren-Smith et al. 1984,
Schl\"otelburg et al. 1988, Fraix-Burnet et al.  1989), with some
regions approaching the maximum possible for optically thin
synchrotron radiation ($\sim 72\%$) both in radio (Owen, Hardee \&
Cornwell 1989, hereafter OHC89) and optical/UV (this paper, Capetti et
al. 1997, Thomson et al. 1995) images.  Such high polarization
suggests the presence of highly ordered magnetic fields in the jet.
Recent VLA observations suggest that the kpc-scale lobes are also
highly polarized (Zhou 1998).

High-resolution radio observations with the VLA (Biretta, Zhou \& Owen
1995; hereafter B95) have detected superluminal motions within the jet
of M87 at distances $\sim$ 200 pc from the nucleus, and speeds $\sim
c$ up to 1200 pc from the nucleus.  At smaller scales,
post-refurbishment {\it HST} +FOC observations in the near-UV have
detected a number of new superluminal components (Biretta et al. 1998,
1999; hereafter B98, B99), with speeds ranging from 2-6$c$.  It is
therefore interesting that VLBI observations have yet to discover any
evidence of superluminal motions on parsec scales (Biretta \& Junor
1995, Reid et al. 1989).  This may be due either to insufficient
temporal sampling, or to differences between the radio and optical
structure of the jet.  A VLBI monitoring campaign with rapid sampling
is now underway to study this issue.

Here we present the results of HST/WFPC-2 optical, and VLA radio,
polarimetry of the M87 jet.  Both datasets were obtained during
1994-1995, and have resolution $\sim 0.2''$, translating to 15 pc
linear resolution at a distance of 16 Mpc.  The optical data we
present are much higher signal-to-noise than pre-COSTAR FOC and WFPC-1
polarimetry data (Thomson et al. 1995, Capetti et al. 1997), and
therefore are particularly useful in revealing many new details
regarding the magnetic field structure of the inner jet which were
only hinted at in earlier data.  They also do not suffer from the
spherically aberrated PSF of pre-COSTAR observations.  They are,
however, somewhat lower spatial resolution, due to the instrumental
configuration used.  Another difference between this study and
previous works (Thomson et al. 1995, Capetti et al. 1997) is that the
radio map to which we compare our results was taken within 15 months
of the optical observations.  This helps to resolve some of the
ambiguities noted in those papers.

In \S 2, we will discuss the observational setup and data reduction
procedures.  Our results will be presented in \S\S 3 and 4, and the
physical implications discussed in \S 5.  In \S 6, we conclude with
remarks centering on the possible impact of future observations.

\section {Observational Setup and Data Reduction}

\subsection{HST Optical Polarimetry}

Optical polarimetry of the M87 jet was obtained May 27, 1995, with the
HST using WFPC2 and the F555W (broadband V) filter plus the POLQ
polarization quad filter.  The observational characteristics of the
instrument are described by Biretta (1996).  Essentially, WFPC2 is an
imager with 4 chips arranged in a chevron.  The PC1 chip has
$0.0455''$ pixels, while the three WF chips (WF2, WF3, and WF4) have
$0.09965''$ pixels.

In order to maximize the unvignetted field of view available for
polarization observations, we used the WF chips, rather than the PC,
for the polarization observations.  For the same reason, we did not
rotate the POLQ quad between observations.  To obtain the Stokes
parameters it was then necessary to obtain images with all three WF
chips, with each one representing an observation through a polaroid
nominally oriented at $PA = 45, 90, 135^\circ - V3$, where $V3$ is the
rotation of the HST's z-axis with respect to the sky (for details, see
Biretta \& McMaster 1997).  A total of 1800 seconds integration time,
split in three to reject cosmic rays, were obtained in each WF chip.
Within the same block of orbits, higher-resolution PC images (2400
secnods integration) were obtained in the F555W and F814W bands.
These data will be discussed in detail in a later paper (Perlman et
al. 1999); however, the F555W PC image is presented in Figure 1a for
reference, and we discuss it briefly in various parts of \S 3 because
its resolution is $\sim$ factor 2 better than possible with the WF
chips only and thus shows considerably more detail in compact regions.

The results of our optical and VLA polarimetry are shown in Figure 1.
That figure includes four panels, which are, respectively, false-color
representations of the optical total intensity and fractional
polarization images; followed by the equivalent in radio.  In all four
panels, the same physical scale has been adopted, and the image has
been rotated so that the jet is along the x axis.  We have marked on
Figure 1 the location and traditional nomenclature for each of the
major knots.  In aligning the HST and VLA datasets, the nucleus was
used as the fiducial point.  The accuracy in the optical/radio
registration should be $\approx 0.02''$, i.e. $\sim 1/5$ of a pixel,
and is limited by the fact that the nucleus saturated all optical
images taken.  In Figure 2, we show a map of the polarization ratio
$P_{opt}/P_{rad}$ for the jet.  The angular scale used in this image
is the same as used in Figure 1, so that the results can be directly
compared.  In Figures 3-6, we show optical and radio contour maps of
various jet regions with magnetic field polarization vectors
superposed.  The Stokes $I$ map used for those figures was produced
with the WF images, rather than the PC, in order to avoid confusion
due to differing resolutions.

\subsubsection {Basic Data Reduction Techniques}

The HST data were recalibrated using the best available flat fields
and dark count images within STSDAS using standard techniques
(Holtzman et al. 1995).  Once this was done, the individual images
were examined; it was found that due to pointing jitter, one of the
three WF3 images was shifted by approximately 2 pixels, and so had to
be shifted back.  The data were then combined and cosmic rays rejected
using the task CRREJECT.  Once the data were flat-fielded and
dark-count subtracted, the three images were geometrically rectified
using the IRAF task WMOSAIC.  After that, an 800 $\times$ 800 pixel
section was cut out of each image, with the nucleus of M87 roughly at
the center.

These sub-images were then rotated so that North was along the y axis,
and then galaxy-subtracted.  Galaxy subtraction was done using the
tasks ELLIPSE, BMODEL and IMCALC in STSDAS. To successfully model the
galaxy, it was necessary to mask out the jet and SE hotspot, as well
as all globular clusters and stars in the field.  This is of necessity
an iterative process, as the fainter globular clusters do not become
apparent until galaxy subtraction is done.  Failure to subtract a
globular cluster would produce a circular ``ringing'' centered at the
distance of the cluster.

Once this was done, the resultant images were compared in detail.  By
measuring the FWHM of globular clusters which were essentially
unresolved by the WFPC observations, it was determined that the WF2
observation had a PSF marginally narrower than the other two
observations (2.05 pixels compared to 2.3 pixels), due to the
redistribution of flux which occurs during the interpolations done by
WMOSAIC and IMLINTRAN (both of which assumed the WF2 images as
fiducial).  To rectify this, the WF2 image was convolved with a
circular Gaussian of $0.13''$ FWHM.  As a result of this procedure, we
were able to determine that maximum resolution of these optical
observations is $0.23''$ FWHM.

\subsubsection{Polarization Calibration and Registration}

The most difficult part of the data reduction process was the
astrometry required to combine the WF2, WF3 and WF4 images into Stokes
{\it I, Q} and {\it U}.  High astrometric accuracy was required to
eliminate residual errors.  Several different methods were tried,
including solutions with up to 40 globular clusters using several IRAF
tasks, and cross-correlation procedures.  Even after the best fit from
these procedures was produced, some additional tweaking was required
(probably due to the undersampled PSF); this was done by hand by
minimizing the ``polarized'' light from galactic regions in non-galaxy
subtracted images (checks using the polarization observed from
globular clusters were also done).

To test not only the robustness of the registration over the entire
image, we gradually ``de-registered'' both galaxy-subtracted and
non-galaxy-subtracted images (along several directions), and then
combined them according to the prescription given by the WFPC
Polarization Calibration Tool.  The results of these tests confirmed
that the WF images are correctly registered to within $\pm 0.15$
pixel.  Visual inspections of globular clusters near the edges of the
800 $\times$ 800 pixel image sections ({\it i.e.}  images which were
run through the polarization calibration tool) were done, to test the
robustness of the registration in rotation.  No shifts could be seen.

As a test of our polarimetric calibration we used aperture photometry
to measure the fractional polarization of both the galaxy light and
bright globular clusters in the image.  All the results were
consistent (within the uncertainties) with unpolarized emission, as
one would expect.

Once properly registered, the three galaxy-subtracted WF images were
combined using the WFPC2 Polarization Calibrator tool (described in
Biretta \& McMaster 1997) to produce Stokes {\it I, Q} and {\it U}
images.  The tool takes into account not only the orientation of the
HST, but also the instrumental polarization terms for all optical
elements of the HST + WFPC2, including the pick-off mirror and the
polarizer filter (which has a high cross-polarization transmission).
Stokes {\it Q} and {\it U} images were then combined in a standard way
to produce percent polarization (defined as $P = (Q^2 + U^2)^{1/2}/I$)
and magnetic field position angle (defined as $MFPA = {1 \over 2}
\times \tan^{-1}(U/Q) + 90^\circ$) images.

The resultant $P$ image is shown in Figure 1b.  This image has been
clipped so that the fractional polarization is shown only for points
with $>100$ ADU above galaxy.

\subsubsection {Uncertainties}

Due to the fact that the Stokes parameter images are each formed by
linear combinations of the three galaxy-subtracted WF images, the
Poisson uncertainty in $P$ at any given point in the jet is
approximately equal to the sum in quadrature of the Poisson errors in
the counts on-jet of any given feature (in each of the three WF
images), plus the uncertainty due to subtracting the galaxy.
Extensive testing (described below) determined that a third factor
also contributes to the uncertainties in our data; this third factor
is the uncertainty in registering the WF images.  Here we discuss the
relative impact of each of these uncertainties on our results.

The Poisson errors due to galaxy subtraction and photon noise vary
widely from point to point along the jet, both as a result of the
large range of surface brightness of jet features, as well as the
relatively steep gradient and large signal from the galaxy.  The
signal from the galaxy (in Stokes $I$) varies from $\sim$ 2600
ADU/pixel $\sim 1''$ from the nucleus, to $\sim 300$ ADU/pixel $\sim
25''$ from the nucleus.  By comparison, the signal from the jet varies
from only a few dozen ADU/pix in some of the faint inter-knot regions
of the inner jet, to nearly 6000 ADU/pix in the brightest part of knot
A.  As a result, the dominant source of Poisson error is different in
various regions of the jet.  In the brightest regions of A, for
example, the jet is far brighter than the galaxy, so that the Poisson
error in $P$ is dominated by the jet signal and is $\sim 1.5\%$ ({\it
n.b.} the WFPC2 gain is 7 electrons per ADU).  A similar situation
pertains in lower surface brightness regions of A as well as B, where
more typical values for the signal are $\sim 1000-2000$ ADU/pix
compared to $\sim 500-700$ ADU/pix from the galaxy.  Poisson errors in
these regions are more typically $\sim 3\%$.  In the faintest regions
of the outer jet, however (i.e., the inter-knot regions between knots
B and C, or low-surface brightness regions in the downstream part of
knot C), the galaxy is much brighter than the jet and as a result the
two contribute much more equally to the uncertainty.  For example, in
the inter-knot region between B and C, the galaxy signal is about 500
ADU/pix while only $\sim 50$ ADU/pix of jet emission is detected in
each WF image ({\it i.e.}, $I \approx 150$ ADU/pix), so that the
Poisson error due to galaxy subtraction is 5.6\% and that due to the
jet signal itself is 9.3\%.  Interior to knot A, the galaxy is
brighter than the jet at every point.  Nevertheless, the relative
contribution of galaxy subtraction and jet to the Poisson errors still
varies considerably.  In the brightest parts of D (the brightest
feature in the inner jet), for example, the signal from the jet is
$\sim 500$ ADU/pix in each WF image, compared to about 2000 ADU/pix
(in Stokes $I$) from the galaxy, so that galaxy subtraction
contributes a Poisson error of 1.1\% and the jet contributes a Poisson
error of 2.9\%.  The balance is somewhat different in lower
surface-brightness regions.  For example, in the inter-knot regions
between D and E, the surface brightness of the jet is 100-150 ADU/pix
(in $I$), whereas the galaxy at that point has a surface brightness of
$\sim 1500$ ADU/pix.  In these regions, the Poisson uncertainties from
galaxy subtraction and the jet signal are about equal, and each is
$\sim 10\%$.  It is important to note that only in the lowest
surface-brightness regions do these two factors contribute about
equally to the errors.

In most regions of the jet, the dominant contribution to our
uncertainties comes from the registration of the three WF images, and
not Poisson noise.  The de-registration procedure described in \S
2.1.2 also helped us establish the magnitude of this uncertainty,
since one can closely examine the {\it I, Q, U, P} and MFPA images
thus produced.  The effect of a $\pm$ 0.15 pixel de-registration along
the direction of the jet upon the fractional polarization is $\delta P
\lsim 0.05$ (not 5\%) in most jet regions, and $\delta P=0.1$ in the
region of knot A, where the intensity gradient is by far the largest
of any region in the jet.  The position angle is somewhat more stable:
the effect of a $\pm$0.15 pixel de-registration along the direction of
the jet upon the position angle of polarization was $\lsim 5^\circ$
throughout the jet.  That this dominates the uncertainties in our
result can be seen by comparing the size of this effect with the above
figures for Poisson uncertainty, taking into account that the values
we see for $P$ range from 0.03 to $\sim 0.6$ in the most heavily
polarized regions.  Only in the faintest regions of the inner jet do
Poisson errors dominate.

The errors we quote herein on the optical polarization of individual
resolution elements will include all of the above-mentioned effects
summed in quadrature.  When examining larger regions, scatter will
also be taken into account.

\subsection{VLA Radio Polarimetry}

Radio data were obtained at the VLA on February 4, 1994 using the A
configuration.  Only one IF (14.435 GHz) was used, due to polarization
calibration problems in the data in the other IF.  The continuum
correlator was used for this experiment, giving us a maximum bandwidth
of 50 MHz.  The total integration time was 10 hours.  The observations
also included a nearby phase calibrator, the flux and polarization
calibrator 3C 286, and the closure error calibrator 3C 273.  The
observations were made with the phase center south of the jet, at RA
(1950) = 12 28 17.000, DEC (1950) = 12 40 00.00.

The hybrid mapping process was started using CLEAN components from a
previous observation in January 1993 at the same frequency (B95).
This initial cross-calibration serves to minimize differences in the
data sets which might arise from calibration errors.  Subsequent
iterations of self-calibration should then remove any constraints
incorrectly imposed by the cross-calibration.  Repeated iterations of
MX and CALIB were done until a final map was obtained (see B95 and
Zhou 1998 for further details)

At this point we used the bright point source in 3C273 to calibrate
the instrumental polarizations due to the antennas and feeds with PCAL
in AIPS, and used the 3C286 data to calibrate the polarization
position angles with CLCOR.  After these solutions were applied, a
final run of CLEAN was done.  The final CLEAN image had $0.15''$
resolution and $0.045734''$ pixels (both identical to the image
presented in OHC89).  For direct comparison with the HST image, we
have smoothed the radio image to $0.23''$ resolution and resampled to
$0.09965''$ pixels.  This improved the signal-to-noise, which was
important in lower surface-brightness regions. We show this image in
Figures 1c and 1d.

\section {Optical Polarization Structure}

As can be seen in Figure 1, the jet is highly polarized over nearly
all of its length.  In the knot regions, $<P_{opt}> \sim 0.3$.  Many
of the knots have sections which are $\sim 40-50\%$ polarized,
attesting to the highly ordered magnetic field structure which
persists for well over 1 kpc from the nucleus.  The most highly
polarized regions of the jet are in knots HST-1, D, A and B, which
includes the brightest knots as well as those where superluminal
motion has been found (B95, B99, Zhou 1998). Optical polarization
parameters could not be computed for the nucleus, as the central 4
pixels were saturated.  The optical MFPA direction in the inner jet
(interior to knot A) is predominantly along the direction of the jet.
However, close inspection reveals complex structure in each knot.  The
optical MFPA structure observed in the outer jet knots (A, B, C and G)
is very complex.

No polarized emissions are detected in the SE hotspot.  However, this
does not contradict previous results ($P=0.34 \pm 0.05$; Sparks et
al. 1992), as the surface brightness of the SE hotspot is only $30-35$
ADU/pix at maximum on our $I$ image, compared to a signal of $\sim
350$ ADU/pix from the galaxy ({\it i.e.}, 1 $\sigma$ uncertainty in
$P$ of 0.33).  We did attempt to search for compact, highly polarized
emissions in this region by smoothing our Stokes $I$, $Q$ and $U$
images with a Gaussian of 5 pixel FWHM.  This improved the 1$\sigma$
uncertainty to $\delta P = 0.20$ but nevertheless no compact, highly
polarized regions were detected.

Here we discuss our optical polarimetry results in the light of the
optical morphology.  We divide the discussion into subsections on the
inner (interior to knot A) and outer jet, respectively, since as has
been noted by many authors (e.g., SBM96 and references therein), the
character of the jet changes drastically at knot A.

\subsection{The Inner Jet}

The morphology of the inner jet in both the optical and radio is
dominated by five bright knot regions, usually termed HST-1 ($1''$
from the nucleus), D ($2.5-4''$ from the nucleus), E ($6''$ out), F
($8''$ out) and I ($11''$ out).  Significant inter-knot emission is
observed, although, as SBM96 have noted, the knot-to-inter-knot
contrast is much higher in the optical than in the radio.  We do
detect polarized optical emissions in some of the inter-knot regions,
but at relatively low significance ($\sim 2-3$ sigma).  In the ensuing
discussion, we will look at the characteristics we observe in each of
the bright knots.

In the PC image shown in Figure 1a, knot HST-1 is characterized by a
bright knot and fainter emission trailing downstream from the bright
knot.  Even higher resolution imaging and yearly monitoring by HST
(B98, B99) resolves the fainter emission into several components, each
of which appear to be moving superluminally with speeds up to $6c$.
Variability on 1-year timescales is also observed in these knots (B98,
B99).  The resolution of our WF data (Figure 3a) is not adequate to
resolve individual, superluminally moving components.  However, we
clearly observe considerable structure in the first $2''$ of the jet.
These include low surface-brightness emission extending $\sim 0.5''$
from the nucleus, a bright knot at the position of the superluminal
components ($\sim 1''$ out), and fainter emission further downstream.
Inspection of Figure 3 also reveals some radio-optical morphological
differences.  In particular, in HST-1, the component at the upstream
edge is brighter (relative to the rest of the knot) in the optical
than in the radio, whereas downstream emission is brighter in the
radio than in the optical.  The MFPA appears parallel to the jet
direction over the first $0.5''$ downstream from the nucleus; however,
the surface brightness of the jet is low in this region and galaxy
subtraction is almost certainly a dominant source of errors, as the
galaxy gradient is extremely steep within $0.5''$ of the nucleus.
Immediately upstream of the HST-1 flux maximum, we observe $<P_{opt}>
= 0.45 \pm 0.08$ and an MFPA orientation approximately perpendicular
to the jet direction.  Near the flux maximum, however, $P_{opt}$
reaches a minimum of $0.14 \pm 0.05$ (averaged over a $2 \times 2$
pixel resolution element), {\it i.e.}, we detect of polarization at
the flux maximum at only the $\sim 2.8 \sigma$ level.  Polarizations
averaging $P_{opt} = 0.41 \pm 0.08$ appear to be present at the north
and south edges of the flux maximum region; however, we caution that
our resolution is inadequate to resolve the width of this component so
the origin of these figures must await higher resolution data.
Immediately downstream from the flux maximum, $P_{opt}$ increases to
$0.45 \pm 0.08$.  The MFPA becomes parallel to the jet about $0.3''$
downstream from the optical flux maximum in HST-1.  In \S 5.1.1 we
discuss this constellation of properties (which also occur in knots D,
E and F).  The extremely high polarizations which appear to be present
about $0.5''$ from the nucleus are greatly affected by galaxy
subtraction (the surface brightness of the galaxy at this point is
$\sim 3500$ ADU/pix while the jet is only $\sim 150$ ADU/pix in $I$).

We turn next to knot D, another highly active region, in which our HST
monitoring (B98,B99) has revealed several components moving at
apparent speeds up to $5c$.  Our data (Figure 3a) reveals significant
polarized structure in the knot D complex.  At the upstream end of
D-East, $<P_{opt}> = 0.30 \pm 0.06$, and the optical MFPA is
perpendicular to the jet.  Similarly to knot HST-1, we detect very
little polarized optical emission ($P_{opt} < 0.1$ is the $2\sigma$
upper limit) at the position of the flux peak in D-East, and the
polarization remains low in other interior regions of D-East.
However, along the northern and southern edges of D-East, significant
polarization is seen, with $P_{opt}$ varying from 0.15 to about 0.60
(typical $1 \sigma$ error = 0.06) and the optical MFPA is primarily
parallel to the jet axis.  There is a peak at the southern edge of
D-Middle where $P_{opt}$ reaches 0.85; this value is likely affected
by small number statistics and interpolation as this is a low surface
brightness region of the jet.  Just upstream from the flux maximum in
D-West, the optical MFPA begins to rotate both at the edges and in the
center of the jet; at the position of the flux maximum in D-West the
optical MFPA is rotated by $\sim 45^\circ$ and appears to remain
roughly constant through the remainder of the knot.
  
Knot E, $\sim 6''$ from the nucleus (Figure 4a) is the faintest knot
in the inner jet. The knot is characterized by lower surface
brightness emission which begins about $1''$ downstream from the
western edge of knot D.  The region of the flux maximum is unresolved
in our WF observations (Figure 4), but at the higher resolution of the
PC image (Figure 1) it is resolved into a double structure.  The
optical and radio emission (Figure 3) are distributed somewhat
differently, with the radio image showing a somewhat brighter eastern
edge.  Because of the low surface brightness of this feature, it is
difficult to determine an average polarization figure for it.  We can
therefore only describe its polarimetric characteristics in more
general terms.  A few tenths of an arcsecond upstream from the flux
maximum, the polarization averages $<P_{opt}> = 0.20 \pm 0.05$ with a
generally east-west MFPA.  Near the optical flux maximum, however, we
do not detect appreciable polarization ($P_{opt} < 0.1$ at $2
\sigma$).  Downstream from the flux maximum, we observe $<P_{opt}> =
0.15 \pm 0.05$ and an MFPA within $\sim 20^\circ$ of the jet
direction.  The data are, however, quite noisy in this region.

Knot F is significantly but not completely resolved at the
$0.2''$(Figure 4a) resolution of the WF images.  At higher resolutions
(Figure 1a), it shows a diffuse double structure, with the brighter
region further from the nucleus.  The polarized fraction at the
upstream edge of the knot F complex is modest ($P_{opt} = 0.15 \pm
0.05$) and barely at the $3 \sigma$ level, but we consider its
detection significant because of the consistency of the MFPA observed
across this region, where it is perpendicular to the jet.  Near the
position of the first flux peak, no polarization is detected (as in D
and HST-1); however, at the position of flux maximum (the second peak
in Figure 1a), $P_{opt}$ has reached $0.20 \pm 0.05$ (the scatter on
individual points is lower), and increases steadily in the downstream
part of the knot complex, reaching a maximum at the northwest edge of
knot F, where an average polarization of $P_{opt}=0.36 \pm 0.06$, and
peaks in individual pixels of $P_{opt} \approx 0.45$, are seen.
Similarly to knot D, the optical MFPA becomes perpendicular to the jet
direction at the upstream edge of the knot.  Along the northern and
southern edges, the optical MFPA is initially parallel to the jet, but
rotates slightly about $\sim 0.5''$ downstream of the head of the knot
to become very nearly east-west.  The mostly east-west {\bf B}-field
persists until the trailing edge of the knot region, where the MFPA
rotates by about $45^\circ$.

Knot I (Figure 5), $\sim 11''$ from the nucleus, is modestly polarized
($<P_{opt}> \approx 0.15 \pm 0.05$).  Slightly higher values of
$P_{opt}$ are observed along its northern and southern edges ($0.20
\pm 0.05$ and $0.26 \pm 0.05$ respectively; an unresolved peak of
$0.41 \pm 0.07$ is seen near the southwestern edge).  About $0.3''$
upstream of the optical peak, we see a rotation by $20^\circ$ in the
optical MFPA, to a nearly east-west orientation; except at the
northern edge of the knot where it is well aligned with the jet
direction.

\subsection{The Outer Jet}

Knot A (Figure 5a) has a highly complex structure, dominated by a
bright, broad shock.  Over most of this region, and especially along
the jet axis, the {\bf B}-field is parallel to the shock front, which
is inclined $18^\circ$ to the jet normal.  Typically $P_{opt} = 0.35
\pm 0.03$ (uncertainty determined by scatter only, due to the size of
the region; the single pixel maximum is $P_{opt} = 0.47 \pm 0.10$) in
the brightest regions.  In fact, the region with {\bf B} parallel to
the shock front starts about $0.7''$ upstream of the flux maximum, and
interestingly, the most highly polarized portion of knot A is $0.5''$
upstream of the optical flux maximum, corresponding to the pre-shock
``bar'' region (using the terminology of B98). Here we observe an
average polarization $P_{opt} = 0.53 \pm 0.07$, peaking at a
single-pixel maximum of 0.59.  While we must stress that our $0.2''$
resolution is not adequate to resolve the bar from the strong gradient
at the upstream edge of A, we find this feature highly interesting.
We will comment further on it in \S 5.2.1.  Here we note only that it
is important for this feature to be confirmed by higher-resolution FOC
observations.  We have scrutinized the pre-COSTAR FOC observations of
Capetti et al. (1997) for information on this region; those authors do
not specifically comment on this region in their text but an
inspection of their maps reveals it to be highly polarized.  We
observe a somewhat lower polarization region ($P_{opt} \sim 0.30$)
separating the ``bar'' region from the flux maximum.  The degree of
polarization in the jet interior decreases gradually downstream of the
bright shock, and the optical MFPA remains roughly aligned with the
shock in this region.  The edges of the jet in knot A have a more
complex behavior than the interior.  Throughout the edges, $P_{opt}$
is more modest than in the brightest regions, averaging $\sim 0.15$.
The MFPA rotates significantly as one moves from the jet center
towards the edges, even at positions equidistant from the nucleus
compared to the flux maximum.  This rotation reaches almost $90^\circ$
({\bf B} parallel to the jet axis) in jet-edge regions about $0.5''$
downstream from the flux maximum; a similar MFPA orientation is not
observed in the jet interior until one moves downstream by another
$0.5''$.
 
Knot B, $14''$ from the nucleus (Figure 5a), is also complex.  It has
two bright regions: B1, which is closer to the nucleus and near the
jet axis, and B2, which is downstream from B1 and at the southern
edge.  Knot B2 is somewhat more polarized than B1, with $P_{opt}$
averaging $0.34 \pm 0.05$.  The most heavily polarized regions of knot
B are at the edges, particularly the southern edge, where $<P_{opt}> =
0.45 \pm 0.05$ and the single-pixel maximum is 0.54.  The interior of
the jet and particularly B1 is only modestly polarized, with $P_{opt}$
averaging $0.22 \pm 0.07$.  But even within B1, the southern part is
more heavily polarized, reaching a maximum of $0.41 \pm 0.05$ near the
southwestern edge of B1.  Within B1, the MFPA is oriented parallel to
the jet direction.  However, $\sim 1''$ downstream of the optical flux
peak of B1, the MFPA becomes perpendicular to the jet.  The MFPA is
parallel to the jet in B2 and also in other lower surface brightness
regions of B.  We will comment further on these properties in \S
5.2.2.

Knot C (Figure 6) is located $17''$ from the nucleus.  It is
characterized by a bright, rhomboidally-shaped region $\sim 1''$ long,
which is wider at the downstream end.  The jet itself flares somewhat
in this region, its width increasing by nearly a factor 2 in about
$2''$.  A second, fainter shock (which can be seen at higher
resolutions, e.g., Figure 1) appears at the downstream edge of the
flaring region.  We observe fairly complex polarization structure in
C.  $P_{opt}$ reaches a minimum all along the eastern and northern
edges of the rhomboidal main shock; these regions can be seen quite
easily on Figure 6 and have $P_{opt} < 0.10$ at $2 \sigma$ (see \S
5.2.2 for further discussion).  The most heavily polarized, bright
part of knot C is just downstream from the flux maximum, where
$P_{opt}$ reaches $0.30 \pm 0.04$ (uncertainty determined by scatter).
Comparably high polarizations (averaging $0.26 \pm 0.05$ and peaking
at $0.36$) are observed at the downstream edge of the knot, where the
MFPA remains perpendicular in the jet interior.  The MFPA assumes a
nearly perpendicular orientation near the upstream edge of the main
shock region, but is parallel to the jet along the northern and
southern edges of the entire knot C region.

Knot G, $20''$ from the nucleus, is very low surface brightness, and
characterized by two diffuse knots, often called G1 and G2 (Figure 6).
We detect fairly high polarization (at relatively low significance)
throughout both G1 and G2, with $<P_{opt}> = 0.30 \pm 0.10$.  The MFPA
appears to be oriented within $30^\circ$ of the jet direction
throughout knot G.

\section {Radio Polarization Structure}

We now turn to the radio data.  Since OHC89 have already given
detailed descriptions of the structure (both total and polarized flux)
in various jet regions, here we discuss only regions where significant
changes are observed.  As the main emphasis of this paper is on the
polarized structure, we will only detail structural changes in general
terms.  For a more detailed analysis of proper motion and variability,
the reader is referred to Zhou (1998).

The radio morphology of the inner $2''$ of the jet has changed
markedly since 1985 (Figure 3b).  The most noticeable change is the
appearance of a bright region which begins $0.8''$ from the nucleus
and extends for nearly $1''$.  This feature, which corresponds roughly
to the location of the HST-1 region, can also be seen on the 1993 Jan
11 map shown in B95.  This change in structure is most likely the
result of motions in the jet, consistent with our findings (B98, B99)
of components with speeds up to $6c$, which vary on 1-year timescales,
in HST-1.  By comparison, the 1985 map showed a relatively bright
region which began $1.1''$ from the nucleus and extended for about
$0.8''$ (OHC89).

The radio MFPA in HST-1 is perpendicular to the jet axis at the
position of the first peak, but within $0.3''$, resumes a generally
parallel orientation.  Interestingly, no significant deviations from a
parallel orientation were seen in the first $2''$ of the jet in the
1985 map (OHC89).  Throughout the HST-1 complex, $P_{rad}$ remains
modest, ranging from 0.13 to 0.33 and averaging 0.21.  This behavior
is significantly different from that observed in the optical and is
dealt with in \S 5.1.1.

A change is also noticeable in the knot D complex (Figure 3b), where
the separation between knots D-East and D-West (measured peak-to-peak)
has changed from $0.76''$ (OHC89) to $1.04''$.  This is very close to
the value of $1.12''$ measured by Capetti et al. (1997) on pre-COSTAR
HST images; we therefore assert that the radio-optical differences in
the structure those authors noticed were largely due to temporal
evolution.  We believe this evolution is likely to include both motion
and variability, rather than just motion (which, if alone responsible,
would imply a speed of $7.9c$).  The peak of emission in D-East
appears to be about $0.2''$ closer to the nucleus in our 1994 image
than in the 1985 image.  In addition, knots D-Middle and D-West are
known to move superluminally with speeds $\sim 2.1-2.5 c$ (B95) in the
radio.  The combination of these two easily explains the change in
D-East to D-West distance we see (see Zhou 1998 for a deeper
analysis).

At the upstream end of D-East, the radio MFPA is approximately
parallel to the jet direction.  A $\sim 20^\circ$ rotation (to nearly
east-west) can be seen beginning near the flux peak of D-East and
continuing through D-Middle.  Significant rotation is seen in D-West,
similar to what is seen in the optical (\S 3.1).  As in HST-1, the
degree of polarization varies little ($<P_{rad}> = 0.32$, maximum
$P_{rad}=0.45$) throughout the knot D complex and does not reach a
minimum near the flux maximum of D-East as observed in the optical
(see \S 5.1.1).

The degree of polarization in F (Figure 4b) in the radio varies
considerably more than in knots HST-1 and D, ranging from 0.11 to
0.50, and averaging 0.26.  There has been very little change in the
overall polarization structure of knot F in the nine years between the
OHC89 map and ours; however, since our signal-to-noise is somewhat
higher, some additional detail is revealed.  In both, the radio MFPA
remains roughly east-west from the flux peak downstream (similar to
the optical).  No evidence for rotation of the radio MFPA can be seen
at the upstream edge of the knot.  At the western edge, our higher S/N
allows to see some evidence for a slight rotation in the MFPA, similar
to what is seen in the optical, but considerably smaller in magnitude.
We also see considerable polarized emission in the inter-knot region
between knots F and I, where the radio MFPA is oriented in the
east-west direction.

\section {Implications}

Polarimetry provides clues to the direction of the magnetic field
within the jet, as well as the character of the plasma flow.  Hence it
is one of the most important diagnostics for deciphering jet
structures.  Comparisons between polarimetry results in different
bands give us important further insights into the internal structure
of the jet.  In particular, since previous HST observations have
already shown that optical emission in the M87 jet is more
concentrated along the jet axis, a comparison of optical and radio
polarimetry results gives us the opportunity to learn about the
magnetic field structure as a function of distance from the jet axis.

\subsection {Physical Processes in the Inner Jet}

The majority of the differences between our optical and radio
polarimetry results occur in the inner jet.  Several properties are
common to more than one jet region.  We will discuss such regions
together.

\subsubsection {Knots HST-1, D, E and F}

Significant differences are present in the optical and radio
polarization structures of knots HST-1, D, E and F (Figures 3, 4).
First, the degree of polarization varies less in the radio than in the
optical.  The average optical polarization just upstream of the flux
peak in HST-1 is $<P_{opt}> = 0.45 \pm 0.08$, while at the maximum,
$P_{opt}$ drops to just $0.14 \pm 0.05$.  Similarly, in D, E and F,
$P_{opt}$ values of $0.15-0.30$ are observed just upstream of the flux
peaks, compared with $<0.10$ at the flux peaks. By comparison,
$P_{rad}$ is much more stable, averaging $0.24 \pm 0.07$ upstream of
the flux peak in HST-1 compared to 0.15 at the flux peak. A similarly
small change is observed in F, where $P_{rad}$ drops from $0.30 \pm
0.09$ to $0.20 \pm 0.07$ at the position of the flux peak, although
{\it n.b.} there is a local minimum of 0.11 $0.2''$ further south.
The drops in $P_{rad}$ observed at the flux peaks of D and E are
statistically insignificant.  At the upstream ends of knots HST-1, D
and F, the optical MFPA rotates to a direction nearly perpendicular to
the jet, whereas the radio MFPA remains primarily along the jet axis
in D and F.  At the downstream edges of these components, the MFPA are
generally more similar in the radio and optical.

M87 is not the only object where high polarization and perpendicular
MFPA is observed near the upstream edge of bright knots, and the
polarization decreases near the flux maximum.  Similar properties are
observed in radio observations of the inner jet of Cen A (Clarke et
al. 1992).  Unfortunately, no three-dimensional view of those
structures is yet available due to the large optical obscuration
present in Cen A.

The differences in the optical and radio polarization characteristics
we observe are clear evidence that the optical and radio emission seen
from the jet originate in somewhat different physical regions.  In
Figure 7, we illustrate a model for the physical situation in these
knots.  The critical features of this model are (1) that the {\bf
B}-field direction varies as a function of distance from the jet axis,
and (2) that the optical and radio synchrotron emitting electron
populations are not completely co-located.  The optical emission is
dominated by the bright, ``shock-like'' regions (likely in or near the
jet center), where {\bf B} is perpendicular to the jet axis due to
compression of the magnetic field.  By comparison, the radio emission
is dominated by less energetic electrons near the jet surface, where
{\bf B} is parallel to the jet axis possibly due to shearing of field
lines against the external medium.  In this model, the $P_{opt}$ in
bright regions will either be low or have MFPA perpendicular to the
jet axis, depending on the relative intensities of the {\bf B}
parallel surface emission and the {\bf B} perpendicular emission near
the jet center.

Taken by themselves, our radio data do not contradict the model
proposed by OHC89 (which was proposed for the 1985 radio data).  The
OHC89 model postulates a surface layer with uniform emissivity $\sim
0.1 \times$ the jet width, a negligible interior emissivity, and a
bright filament wrapped helically around the jet to create the
asymmetry.  However, the OHC89 model cannot explain the differences we
observe between the optical and radio polarization structure in knots
HST-1, D, E and F (above), nor can it explain the subtle but
significant morphological differences ({\it e.g.,} higher
knot-to-interknot region contrast; SBM96) in the inner jet.  Indeed,
it is very difficult if not impossible to explain the differences in
polarization structure under any model where the radio and optical
emitting electron populations are co-located.

It is important to note that our model does not contradict the idea
that Kelvin-Helmholtz instabilities play an important role in
generating many of the features in the M87 jet.  While the energy
input from the Kelvin-Helmholtz mechanism peaks at the jet surface
(Hardee 1983), the pressure perturbation for axisymmetric helical
($n=1$) Kelvin-Helmholtz modes reaches maximum well inside the jet
sheath.  The compression of the magnetic field would be expected to
vary closely with the pressure.  Therefore, we would expect to observe
synchrotron-emitting electrons with the highest energies in interior
regions of the jet where the pressure perturbations reach maximum.

Several groups have explored largely similar models for active jet
regions (Cawthorne \& Wardle 1988; Cawthorne \& Cobb 1990; Clarke \&
Norman 1982; Fraix-Burnet \& Pelletier 1991; G\'omez et al. 1994,
1995a, 1995b, 1997; Hardee \& Norman 1988; Hardee \& Clarke 1995;
Hardee, Clarke \& Rosen 1997; Hughes, Aller \& Aller 1985, 1989, 1991;
Mioduszewski et al. 1997; van Putten 1996).  While they differ in
details, all postulate a shock propagating along the jet, usually with
a propagation speed somewhat slower than the overall flow speed.  In
shocked regions, the magnetic field is compressed, resulting in a more
organized field oriented approximately normal to the jet.  As was
pointed out particularly by G\'omez et al. (1995b), it is by no means
necessary for the shock to encompass the entire jet flow.  The picture
shown in Figure 7 fits quite well with a scenario whereby the shocks
we observe in the inner jet only take up a fraction of the axial
breadth of the jet.  This is also consistent with the observation that
the emission in these knot regions is highly concentrated close to the
jet axis (SBM96).

The wealth of structure we see offers further clues as to the nature
of these shocks.  Quite clearly knots D and F cannot be explained as
simple shocks, given their size, complex structures and velocity
fields (B98, B99).  This is particularly true for knot D, given its
complexity (Figure 1) and the fact that superluminal components have
been found over its entire length (B95, B98, B99).

One interesting possibility is that the shock features may not
necessarily be perpendicular to the overall jet direction.  Two
features suggest this: the decrease in the degree of optical
polarization at the flux peak, and the rise in the degree of
polarization (accompanied by some rotation of the {\bf B-}field) at
the trailing edge of knots HST-1, D and F.  The models of Cawthorne \&
Cobb (1990) and Fraix-Burnet \& Pelletier (1991) allowed the
orientation of planar shock regions to vary (in the other papers
mentioned above, shock regions were assumed to be perpendicular to the
jet direction), and attempted to predict observable features.
Cawthorne \& Cobb (1990) show that under such a model, the degree and
direction of polarization observed in various regions would be highly
geometry-dependent.  A perpendicular MFPA would be observed at both
the upstream and downstream ends.  Along the side edges of such a
shocked region, the observed MFPA would be parallel to the jet
direction.  This model also predicts the observed decrease in
polarization at the flux maxima of these components, since the
position of the flux maximum also represents the position where the
plane of compression of the magnetic field most nearly approaches our
line of sight.  At this position, significant cancellation would be
observed in the projected magnetic field (see G\'omez et al. 1998 and
Cawthorne \& Cobb 1990 for discussions).  We note also that Doppler
boosting could produce brightening by a factor of 10 or more given the
speeds observed in knots HST-1 and D (B98, B99).  It is possible that
the tracks taken by knots within the jet may be curved, perhaps
helical, and that the plane of compression of the magnetic field in
each of these shocks is somewhat inclined with respect to the jet
direction (but is most likely perpendicular to the helical track
followed by the knots).  Shocked regions of the jet could also be the
sites of {\it in situ} particle acceleration, wherein a considerable
hardening of the optical spectrum would be observed ({\it e.g.,}
Begelman \& Kirk 1990).  We have in fact observed this in our
multicolor HST photometry (Perlman et al. 1998).

\subsubsection{Knot I}

In knot I (Figure 5), we see much more similar magnetic field
orientations in the radio and optical; both bands show largely
east-west MFPA and a small decrease in the degree of polarization at
the flux peak.  The situation at the trailing edge of knot I is more
difficult to assess because both the radio and optical maps show
evidence that the polarized emissions from knot A begin only $0.5''$
downstream of the flux peak in I.  These properties do, however,
suggest that the shock which produces knot I is a smaller perturbation
than those which produce the other inner jet knots.

\subsubsection{Inter-knot regions}

The high degree of polarization we observe in inter-knot regions of
the inner jet (typically $20-40\%$; Figure 1) indicates that the
magnetic field in relatively undisturbed regions remains highly
ordered.  The inter-knot regions do not show a significant tendency
for lower polarizations nearer the jet center (though {\it n.b.}  in
our maps the jet is unresolved across its width through the entirety
of the knot D region).  This property is similar to what is observed
in the first $27''$ (12 kpc) of the NGC6251 jet (Perley, Bridle \&
Willis 1984), as well as in the inner few kpc of most other
low-luminosity radio galaxies' jets (e.g., Cen A, Clarke et al. 1992;
3C31; Laing 1998).  Perley et al. (1984) cite this property, as well
as a small expansion rate, as evidence of a high Mach number ($>10$)
in the inner regions of the NGC6251 jet; we suggest that a similarly
high Mach number flow may be present in the inner M87 jet.

\subsection {The Outer Jet}

As many previous authors have noted (SBM96 and references therein),
the appearance of the M87 jet changes drastically at knot A.  From
this point outwards, the jet emission is not so highly concentrated in
a few fairly compact knots.  We observe very high optical and radio
polarizations (close to 50\%) in the outer regions of knot B (Figure
5), as did OHC89.  OHC89 point out that this property is evidence of
lower synchrotron emissivity in the jet core in this region.  An
increase in jet polarization at the outer edges of the jet might also
be evidence of torsional shear, which could not as easily be observed
in the center of the jet because of cancellation of different field
directions seen in projection.  Bicknell \& Begelman (1996) have
modeled knot A as a torsional shock which is eventually responsible
for the disruption of the jet flow observed beyond C.  Our data are
consistent with such a picture.  The torsional shock model is fully
consistent with the structure in the MFPA (which follows the jet
contours at the outer edges), as well as with the suggestions made by
OHC89 regarding the state of matter in the core regions of the outer
jet.  Such features may be common in AGN jets, since similar features
exist in the polarized emissions of at least one other jet (Cen A;
Clarke et al. 1992).

Our data show that the outer and inner jet differ in another way: the
magnetic field structures seen in optical and radio observations of
the outer jet are much more similar than in the inner jet.  In a
global sense, this is an indication that the optical and
radio-emitting electron populations in the outer jet are more closely
co-located than in the inner jet.  Close examination of Figures 5 and
6 does, however reveal a few differences, most of which are also
pointed out as high or low optical polarization regions in Figure 2.
We will detail the implications of these differences below, dealing
with knot A separately, as some controversial claims have been made
regarding its structure.

\subsubsection{Knot A}

High polarization is seen along the edges of knot A (Figure 5), as
well as near the flux peak.  Both the optical and radio maps show
large changes in the MFPA ($\sim 70^\circ$), between knot I and the
bright, broad shock which houses the optical and radio flux peak of
knot A.  The degree of polarization begins to increase about $0.2''$
closer to the nucleus in the optical than in the radio; this can be
seen in our vector map as well as the depolarization image (region
HOP-1; Figures 2, 5).  The rotation of the {\bf B} field vectors also
starts $\sim 0.2''$ earlier in the optical.  Our maps also show some
differences between the optical and radio morphology of A, which were
previously noted by SBM96.  In particular, the upstream end of the
shock region has a concave appearance in the radio which does not
evidence itself in the optical.  While the flux peaks are coincident,
the abrupt brightening at the leading edge of knot A occurs slightly
earlier in the optical than in the radio.  Also, as noted by SBM96,
the jet through this region appears somewhat narrower in the optical
than in the radio, at least at the intermediate and upper contour
levels.

These slight differences between the optical and radio polarimetry
could be explained if the disturbance which illuminates the jet at
knot A begins in the jet interior, and spreads to the entire breadth
of the jet within about $0.3''$ (20 pc).  Under this scenario,
evidence of the disturbance would be seen slightly earlier in the
optical, where higher-energy electrons are concentrated (as indicated
by the inner jet appearing narrower in the optical; see \S 5.1.1 and
SBM96).  The strongly compressed magnetic field in this region could
produce {\it in situ} particle acceleration, which would yield a much
harder spectrum at the upstream edge of knot A.  This might also
explain why the X-ray emission from knot A appears to be located
approximately $0.5''$ short of the optical and radio maximum (Neumann
et al. 1997).  AXAF-HRC observations are required to test this
hypothesis.

Owen \& Biretta (1998) have produced a very high-resolution
($0.035''$) radio map of the M87 jet at 7 mm, with the VLA in A array.
Those data show that the sharp feature at the leading edge of knot A
is narrower than their minimum resolution of $\sim 3$ pc.  One
possible interpretation of the Owen \& Biretta (1998) result might be
that this feature is a filament wrapped around the jet surface, rather
than a shock.  However, three observations reported here and in other
papers argue against this interpretation.  First, the extremely steep
gradient in flux at the leading edge of knot A and the strongly
elevated flux downstream of its flux maximum (relative to the inner
jet), is much more characteristic of a shock than a filamentary
feature.  A second argument which strongly suggests the presence of a
shock within knot A is the marked change in the appearance of the jet
(as several authors have noted); beyond knot A, the jet takes on a
more ``filled'' appearance.  But perhaps the most persuasive argument
of all comes from the polarization vector maps (Figure 5), which show
MFPA parallel to the leading edge of the flux maximum region for more
than $1''$ downstream.  Such a configuration is inconsistent with
identifying the sharp feature with a $\sim 0.03''$ wide filament, but
well in accordance with the presence of a bright shock that affects
the downstream region.

Thomson et al. (1995) claim that there is a region of high
depolarization at the northwestern edge of knot A.  Close examination
of the polarization ratio map (Figure 2) does reveal two small
``depolarization'' features near the western edge of the knot, which
we have labelled HOP-2 and HOP-3; these features have also been noted
on Figure 5.  These features lie within the boundaries of the
depolarization feature claimed by Thomson et al. (1995), but near its
upstream edge.  They are nowhere near as large nor as deep as the
region claimed by Thomson et al. (1995), as close comparison of our
Figure 2 and their Figure 5 shows (we observe maximum depolarization
$\sim 2-3$ compared to values $\gtrsim 5$ reported by Thomson et al.).

Our results do confirm that some ``depolarization'' (low radio
polarization) is present at the downstream edge of knot A.  However,
since our results indicate much smaller differences than did Thomson
et al., we are forced to conclude that their claim of a large mass of
Faraday-active material at the northwestern edge of knot A is probably
in error, and may have been produced by misalignment of the radio and
UV images.  Further evidence for poor radio/UV image alignment can be
seen in their UV-radio spectral index map (their Figure 4), which
appears to show a ``steeper'' spectrum at the upstream ends of knots
F, I and A.  This conflicts with the UV-radio spectral index map made
from higher signal-to-noise (but still pre-COSTAR) data, which
included the nucleus, shown in SBM96.  Such an error may have occurred
because Thomson et al. aligned the radio and UV images by
cross-correlating the optical and radio jet structures.  As has
already been noted, the optical and radio structures are subtly
different; the differences between the two might well produce an
apparent offset of $\sim 0.1''$ towards the Northwest in the HST data.
We also note that comparison of 5 GHz and 8 GHz VLA images (Zhou 1998)
show no excess Faraday rotation or depolarization in this region.

Both ``depolarization'' features are much more consistent with
variations in the energy spectra of emitting electrons as a function
of distance from the jet center, as illustrated for various regions of
the inner jet in \S 5.1.  In particular, the low radio polarization in
HOP-2 and HOP-3 may simply reflect that the radio emission is
returning to its {\bf B} parallel configuration sooner after knot A,
than the optical emission.  This transition region would, of course,
have low polarization, since there is blending of parallel and
perpendicular emission.  That the transition occurs earlier in the
radio may be attributable to shearing at the jet surface, together
with the fact that the jet appears broader in the radio, as noted
before.  The extra component of {\bf B} parallel emission from the jet
surface seen in the radio would tend to make the
perpendicular-parallel transition following knot A, occur earlier.

\subsubsection{Knots B, C and G}

Throughout the remainder of the outer jet, the optical and radio
polarimetry results track each other extremely well.  In both the
radio and the optical data, the highest polarization regions of knot B
are along the edge, and the MFPA is identical.  The story is similar
throughout most of knots C and G.  However, when one looks in more
detail, some small differences appear.

At the northwestern edge of the region between knots A and B, the
degree of polarization remains high in the radio while in the optical,
it becomes nearly undetectably low (this region can also be seen in
the depolarization map; we have labelled it LOP-1).  Immediately
downstream from this region, we see the optical MFPA rotate $90^\circ$
(while the radio MFPA remains approximately east-west), and after
about $0.4''$ it tracks the radio MFPA extremely well.  As can be seen
by inspection of the vector plots, this occurs about $0.8''$ upstream
of the flux maximum in B1.  Capetti et al. (1997) pointed out that the
magnetic field in this region of the jet is highly tangled, and there
is significant structure on scales smaller than $0.2''$, as seen on
their $0.06''$ resolution map (their Figure 5).  When observed at only
$0.2''$ resolution, significant cancellation would take place,
explaining the low optical polarization we observe.

At the northwestern edge of knot B1, the optical polarization
decreases somewhat more rapidly than the radio (but not dramatically
so).  That this is the result of cancellation of magnetic field
vectors can be seen by inspecting Capetti et al.'s $0.06''$ resolution
map, which shows an $0.2-0.3''$ scale magnetic ``loop'' in this
region.  The opposite thing happens near the flux maxima of knots B1
and B2, where the degree of polarization is slightly higher in the
optical than the radio.  In both of these last features, the optical
and radio MFPAs remain consistent with one another.

There is a fairly large region of low radio polarization at the
extreme eastern edge of knot C, which we have labelled HOP-4 (Fig. 2,
6).  This can be seen particularly along the southeastern edge of the
knot, where the optical polarization is somewhat higher and the
magnetic field direction follows the flux contours quite well.
Inspection of Figures 1, 2 and 6 reveals that this occurs just after
the broad minimum reached by both $P_{opt}$ and $P_{rad}$ at the
upstream edge of knot C.  It can be seen by close inspection of Figure
1 that $P_{opt}$ increases slightly quicker than $P_{rad}$, suggesting
that this shock too may begin in the jet interior.  This would require
that (similarly to the situation analyzed above for many inner jet
regions) higher-energy, optical synchrotron-emitting electrons must be
concentrated along the shocked surfaces which are evident as the
bright regions of knot C (particularly at the eastern edge).  Any
particle acceleration would affect higher-energy regions of the jet
first, and hence evidence itself first in the optical.

\section{Conclusions}

The data and analysis we present has given us the first hints
regarding the three-dimensional magnetic structure of extragalactic
jets.  From previous modeling efforts, one would have naively expected
the magnetic fields observed in the optical and radio to be largely
similar, indicating that the high and low energy electron populations
are largely co-spatial.  This was expected particularly because: 1)
the energy input from magnetosonic instabilities (including
Kelvin-Helmholtz) peaks in the jet sheath ({\it e.g.}, Hardee 1983),
and 2) radio observations had suggested that the emissivity in the
center of the M87 jet was much smaller than in its sheath (OHC89).
The present observations reveal a quite different picture: there are
clear differences between the optical and radio polarization
structures of knots HST-1, D, E and F, particularly at their upstream
``shocked'' edges and at their flux maxima.  At the upstream ends of
these regions, we tend to observe perpendicular magnetic fields in the
optical only (in the radio, the MFPA remains parallel to the jet),
while near their flux maxima the degree of optical polarization
declines precipitously.  As previously shown (SBM96), the radio and
optical synchrotron emitting electron populations occupy somewhat
different regions of the jet, with the optical emitting populations
not only more concentrated in the knots, but also closer to the jet
axis.  Hence a natural explanation for the radio/optical polarization
differences we observe, is that the jet contains significant internal
magnetic structure.  Namely, that parallel fields predominate near the
jet surface, while perpendicular fields are present in the knots near
the jet axis, where the field lines are strongly compressed.

Our observations reveal much more similarity in the optical and radio
polarimetry beyond knot A.  This is a strong indication that the
nature of the jet changes drastically at this point, as many other
authors have previously commented.  This leads us to believe that the
processes which create the main emission features in the inner and
outer jet must be somewhat different in nature, with considerable
contributions from shocks in the inner jet, while the the outer jet
more dominated by torsional and/or Kelvin-Helmholtz instability, as
suggested by OHC89 and Bicknell \& Begelman (1996).

The high polarizations that we observe are evidence that even in the
inter-knot regions, the magnetic field remains highly ordered.  The
overall characteristics of the inter-knot regions are somewhat
different in the outer jet than they are in the inner jet.  In the
inner jet, we do not observe a general tendency for the jet center in
these regions to be less polarized than the edges, while beyond knot
A, this characteristic is observed both in the optical and radio.
This is consistent with a high Mach-number flow in inner regions of
the jet, but a somewhat lower Mach number beyond knot A.

Considerable scope remains for future observations of the M87 jet,
based on the HST results pointed out in this and other papers (SBM96;
B98, B99).  This study represents the strongest evidence yet that the
structure of extragalactic jets is stratified in electron energies.
Since the optical-emitting regions seem to be concentrated much closer
to the jet axis than radio-emitting regions (this work), and are more
compact (SBM96), one would expect an increasing compactness at the
highest energies.  The sub-arcsecond resolution of the AXAF HRC will
offer us the opportunity to test this prediction by measuring the
width of the jet beyond knot A as well as pinpoint the location of
knot A's X-ray maximum, which some authors (Neumann et al. 1998)
indicate may be somewhat closer to the nucleus, perhaps in the region
where we see very high optical polarization in these observations.
Similar tests will also be possible in knot B, but not for the inner
jet, which is far too narrow to be resolvable with AXAF.

Further scope also remains for HST observations.  Not only is
significant secular evolution occurring in several regions of the jet
(HST-1 and D in particular; see B98, B99), but also our resolution
($\sim 0.2''$) is insufficient to separate out superluminally-moving
components in these regions.  Higher-resolution observations would
separate individual superluminal components, and more fully resolve
other details of the magnetic field structure where we have noted
these data have insufficient resolution.  It would be particularly
interesting to monitor the evolution of these components as they move
out.  The spectral energy distribution should soften as each component
cools, and harden as particles are accelerated and as new superluminal
components are born.  Moreover, the magnetic field should also respond
to such changes, giving detailed information about the degree of fluid
and magnetic field compression with time (see G\'omez et al. 1995b).
Ongoing and future HST observations should test these possibilities.

\vfill\eject

\vfill\eject

\centerline {\bf{Figure Captions}}

{\bf Figure 1.} False-color representations of the total intensity and
polarization of the M87 Jet in the optical (HST F555W, top two panels)
and radio (VLA 14.5 GHz, bottom two panels).  The HST observations
were carried out in May 1995, while the VLA observations were done in
February 1994.  All maps were rotated so that the jet is along the
x-axis, and are convolved to $0.23''$ resolution.  At the bottom, we
give the false-color scale for the two polarization panels (used for
both optical and radio).

{\bf Figure 2.} A false-color representation of the ratio
$P_{opt}/P_{rad}$ (bottom).  For reference, we also show the optical
intensity map of Figure 1a to the same scale (top).  In the
polarization ratio map, red regions indicate regions where the optical
polarization is higher than the radio polarization, while deep blue
regions are places where the radio polarization is higher.
Considerable differences between optical and radio polarization values
are observed in several areas, which we point out here.  Of particular
interest are the flux maxima of knots HST-1, D-East, E and F, where
the optical polarization decreases markedly (see \S 5.1.1), and
various features in knots A, B, and C (see \S 5.2) which are denoted
HOP or LOP depending on whether the optical polarization is higher or
lower than the radio polarization.

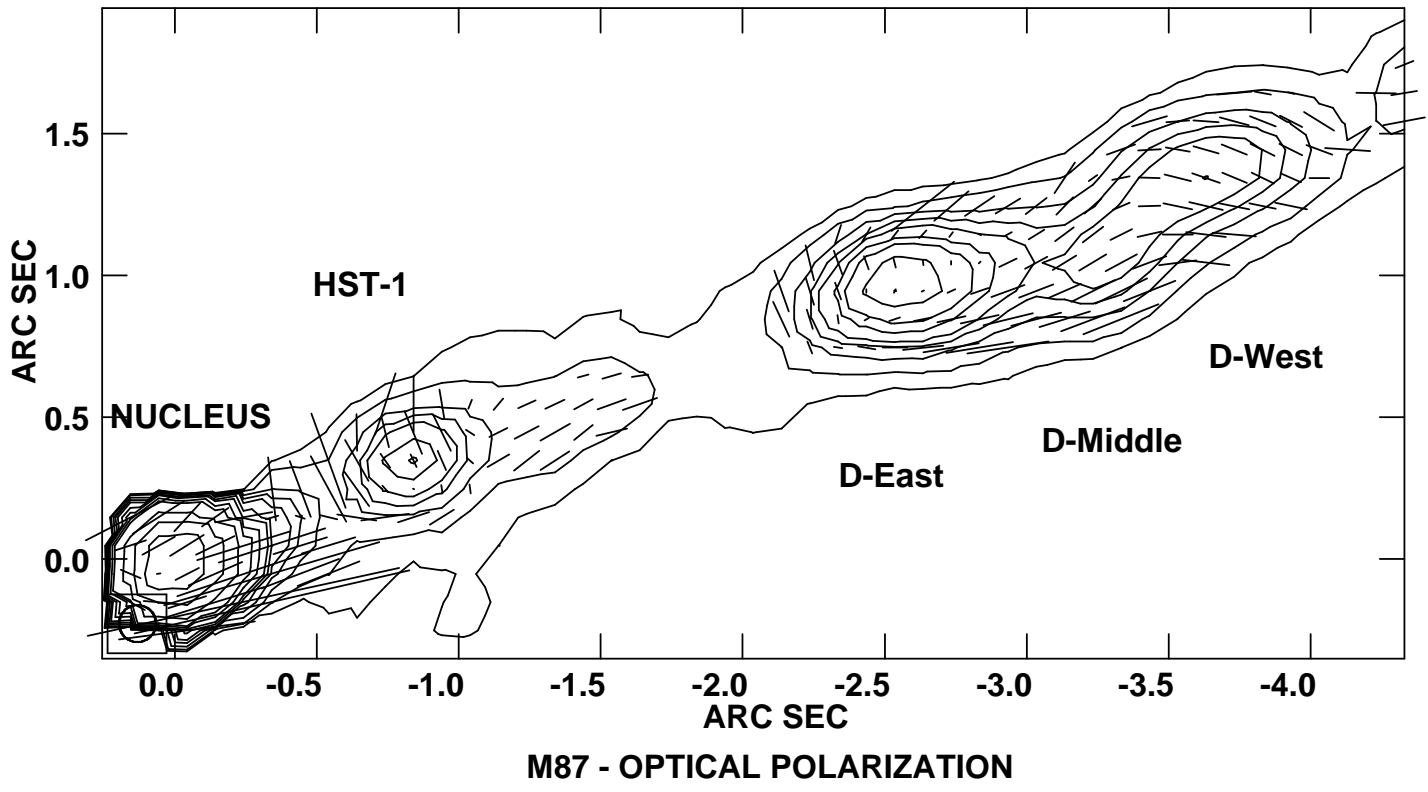
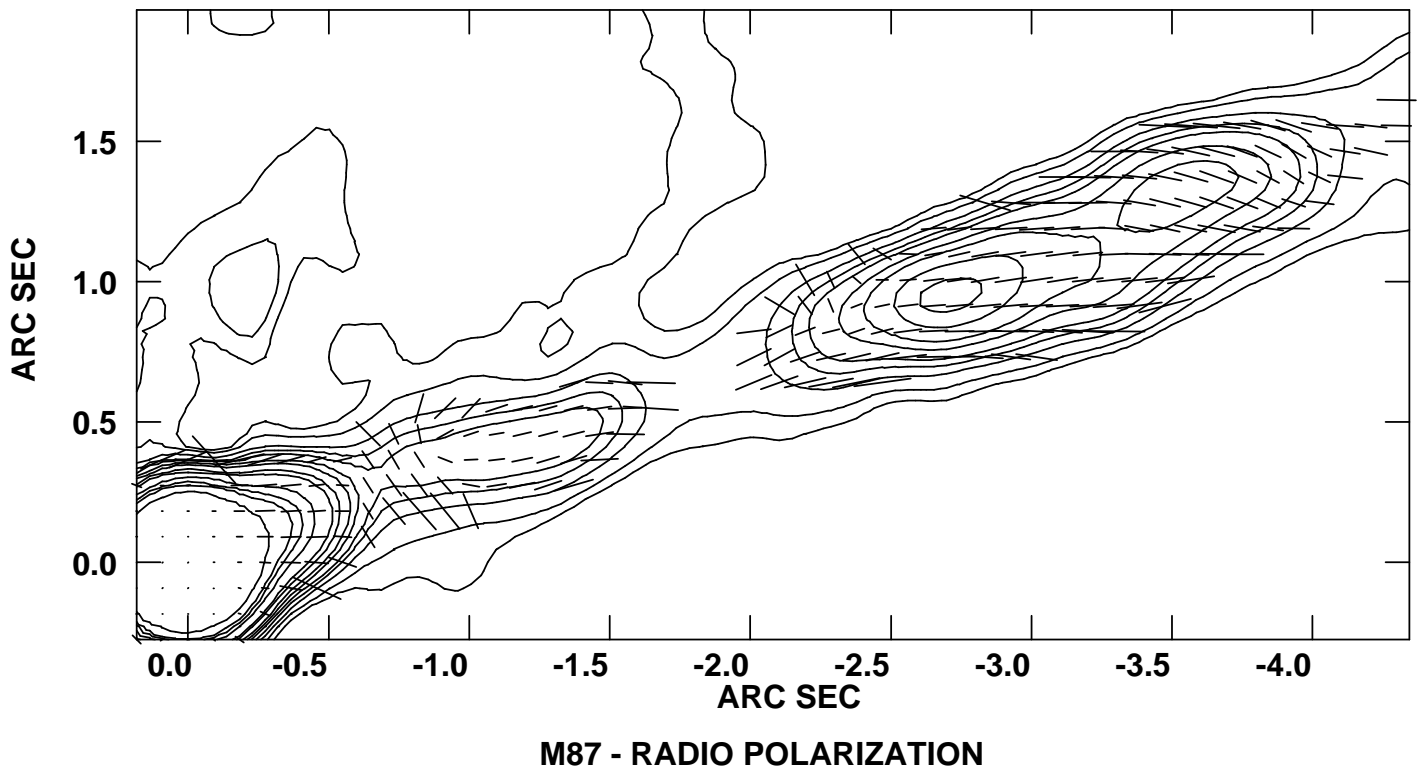
{\bf Figure 3.} At top, we show a contour map of the optical structure
of the innermost jet regions (including the nucleus, HST-1 and D),
with magnetic field polarization vectors overlaid. A 1-arcsecond
vector corresponds to 300\% polarization.  At bottom, we show the
corresponding region in the radio.  North is at the top and East is at
the left.  The optical image is contoured at (1, 2, 4, 6, 8, 12, 16,
24, 32, 64, 128, 256) $\times 50$ ADU/pix, while the radio image is
contoured at (1, 2, 4, 6 ,8, 12, 16, 24, 32, 64, 128, 256) $\times
0.5$ mJy/beam.

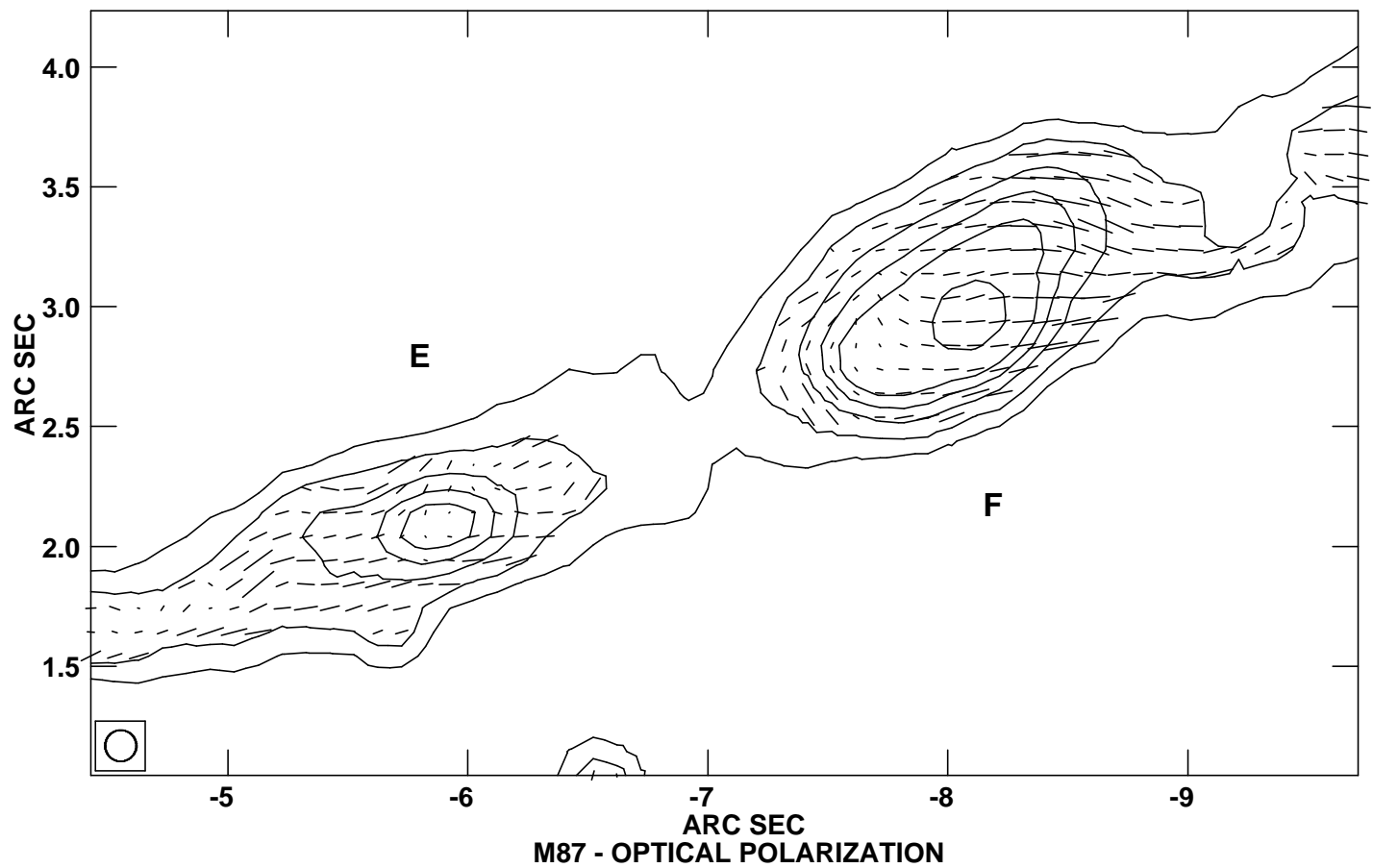
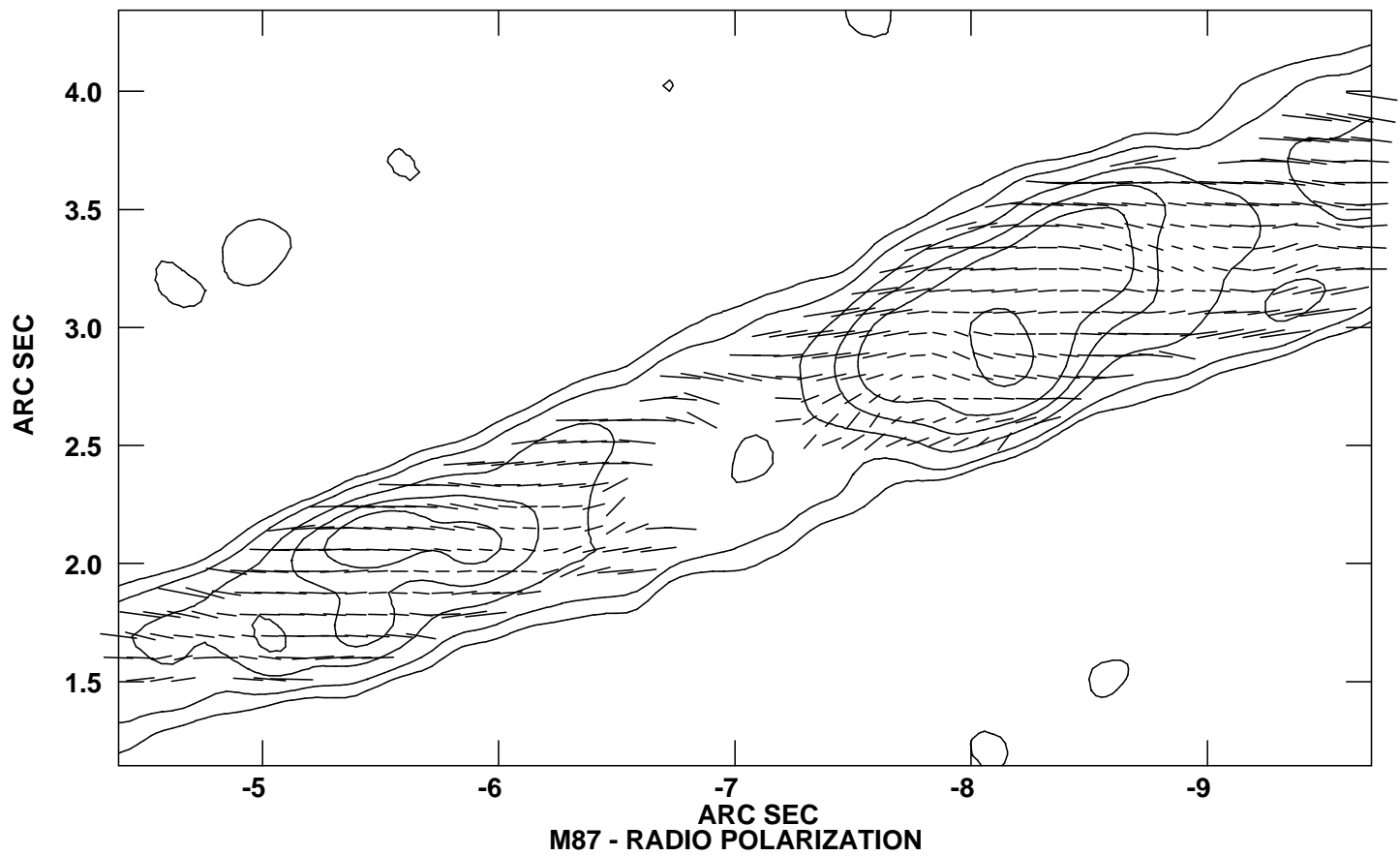
{\bf Figure 4.} At top, we show a contour map of the optical structure
of the knot E and F region of the jet, with magnetic field
polarization vectors overlaid. A 1-arcsecond vector corresponds to
300\% polarization.  At bottom, we show the corresponding region in
the radio.  North is at the top and East is at the left.  The optical
and radio images are contoured as in Figure 3.

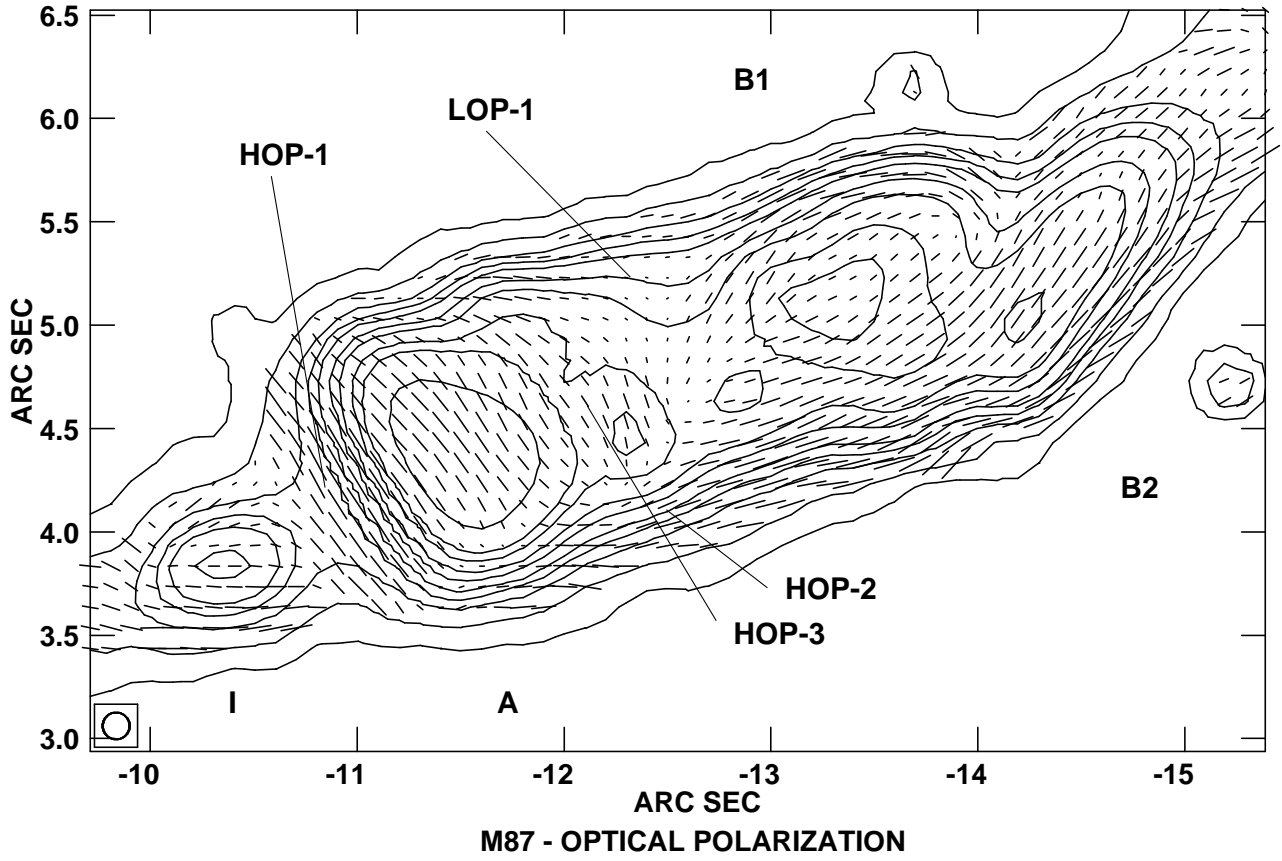
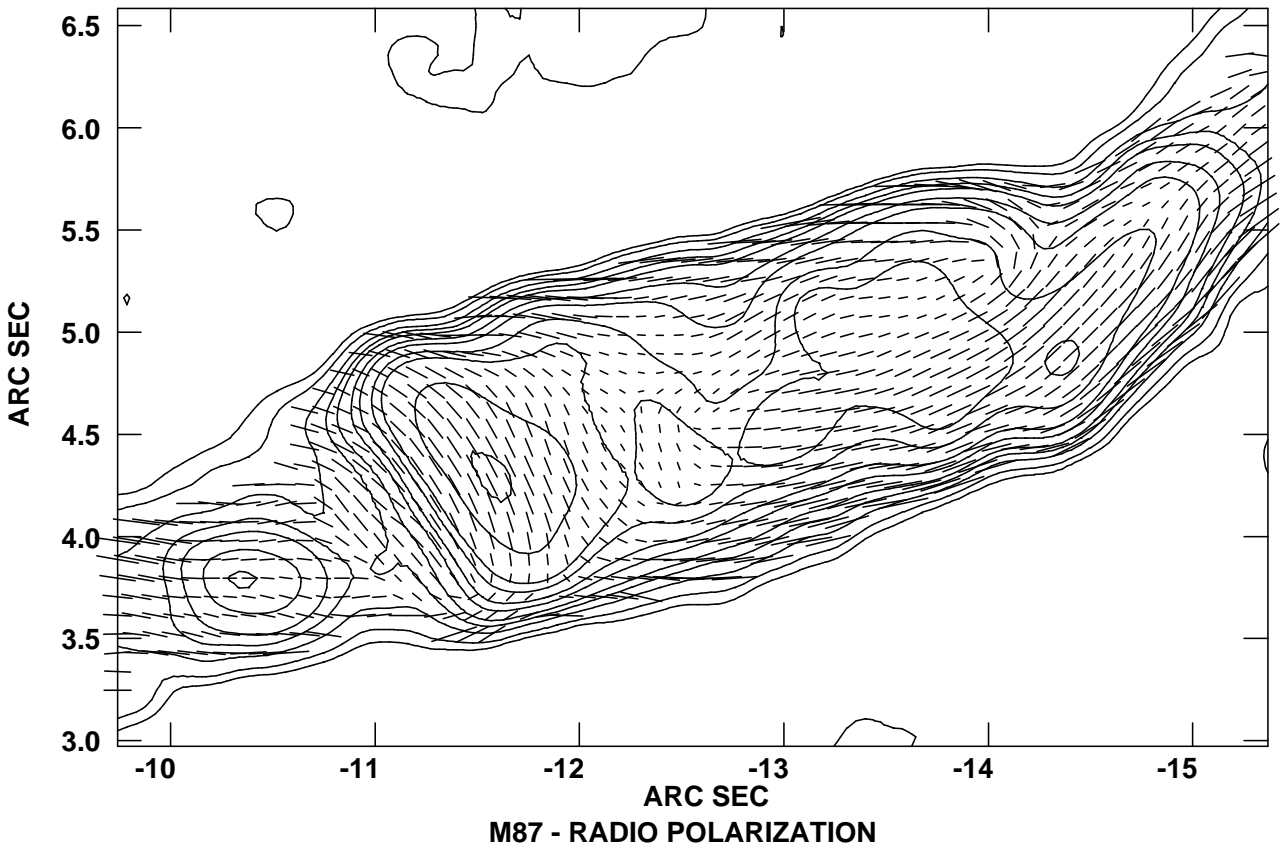
{\bf Figure 5.} At top, we show a contour map of the optical structure
of the knots I, A and B, with magnetic field polarization vectors
overlaid. A 1-arcsecond vector corresponds to 300\% polarization.  At
bottom, we show the corresponding region in the radio. North is at the
top and East is at the left.  The optical and radio images are
contoured as in Figure 3.  The high and low optical polarization
features in this region are labelled to avoid confusion.

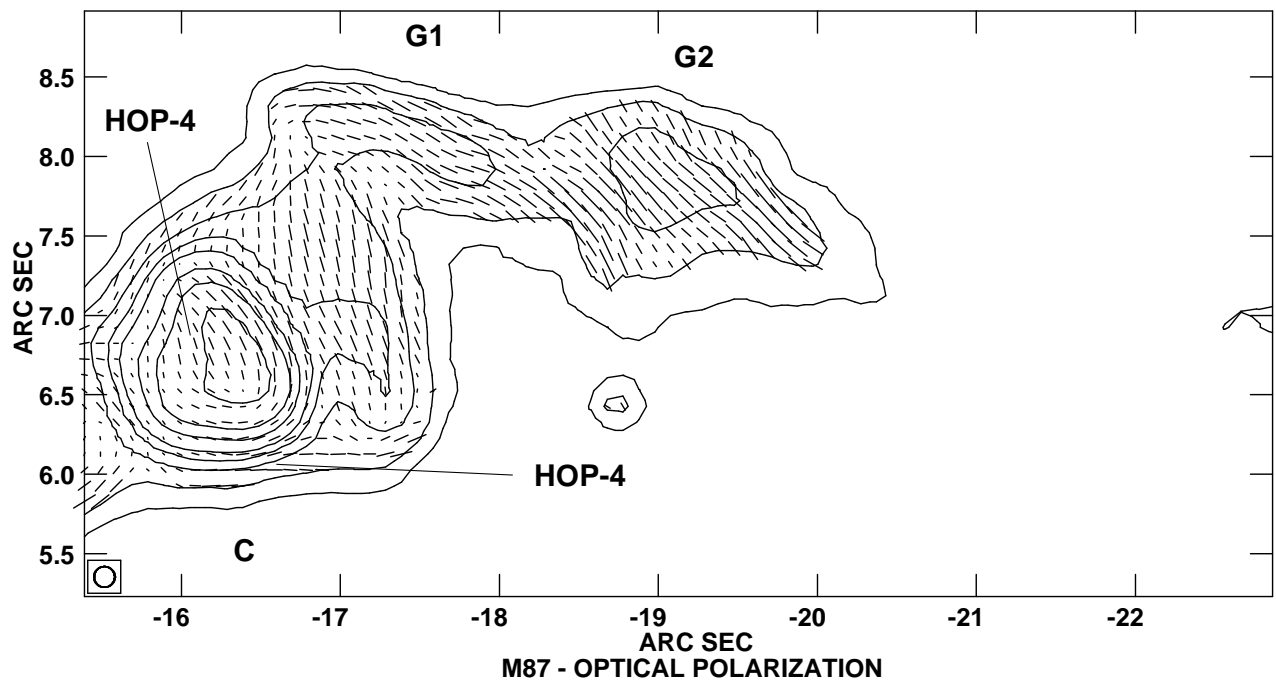
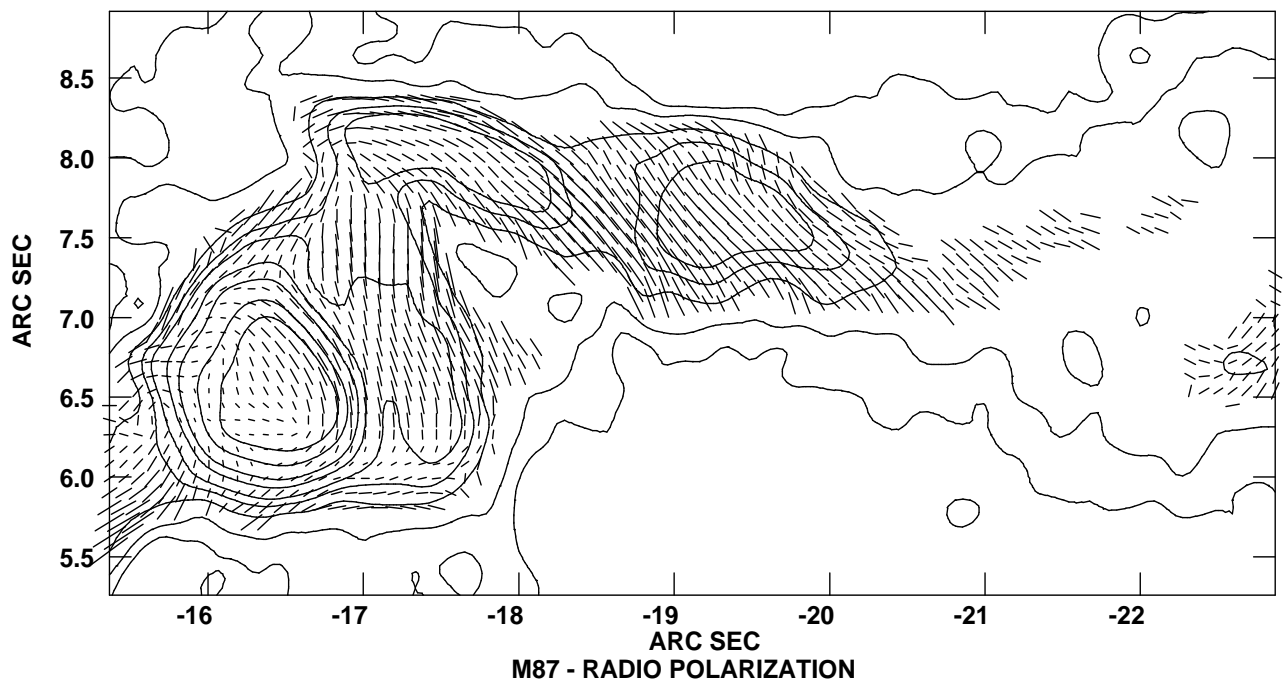
{\bf Figure 6.} At top, we show a contour map of the optical structure
of knots C and G, with magnetic field polarization vectors overlaid. A
1-arcsecond vector corresponds to 300\% polarization.  At bottom, we
show the corresponding region in the radio. North is at the top and
East is at the left.  The optical and radio images are contoured as in
Figure 3.  The high and low optical polarization features in this
region are labelled to avoid confusion.

{\bf Figure 7.} Here we illustrate a model which we believe explains
many of the differences we see both in structure and polarization in
the inner jet.  A key aspect of this model is that optical and radio
emitting electrons are not completely co-located; the optical emitting
electrons are located closer to the jet axis, while most of the radio
emitting electrons are located nearer the jet surface.  Our
polarimetric results suggest significant differences in the magnetic
field structures in these regions.  See \S 5.1.1 for further
discussion.

\end{document}